\documentclass{iau-JDSS}
\usepackage{graphicx}

\title[Connecting reference frames] 
{Connecting terrestrial to celestial reference frames}

\author[Z. Malkin]   
{Z. Malkin\thanks{E-mail: malkin@gao.spb.ru}}

\affiliation{Pulkovo Observatory and St. Petersburg State University, St. Petersburg, Russia}

\pubyear{2013}
\volume{Volume 16}  
\pagerange{1--2}
\date{27 Sep 2012 and in revised form ??}
\jname{Highlights of Astronomy, Volume 16}
\editors{Thierry Montmerle, ed.}

\begin{document}

\maketitle

\begin{abstract}
In this paper we outline several problems related to the realization of the international celestial and terrestrial reference frames --- the ICRF and ITRF ---
at the millimeter level of accuracy, with emphasis on ICRF issues.
We consider here the current status of the ICRF, the interrelationship between the ICRF and ITRF, and considerations for future ICRF realizations.
\end{abstract}

\firstsection 
\vspace{-3ex}              
\section{Introduction}

There are several issues currently preventing the realization of the terrestrial and celestial reference systems (TRF and CRF, respectively) at the mm/$\mu$as level of accuracy:
\begin{itemize}
\itemsep = 0.0ex
\item Insufficient number and non-optimal distribution of active and stable (systematically and physically) stations (VLBI and SLR) and suitable radio sources.
\item Technological (precision) limitations of existing techniques;
\item Incompleteness of the theory/models (e.g., reference systems definition, geophysics); 
\item Not fully understood and agreed-upon details of the processing strategy.
\end{itemize}

The latest ITRF realizations are derived from four space geodesy techniques: VLBI, GPS, SLR, and DORIS, whereas the ICRF is the result only of the global VLBI solution.
The latter is tied to the ITRF datum using an arbitrary set of reference stations.
VLBI also contributes, along with SLR, to the ITRF scale.
Furthermore, all the techniques contribute to the positions and velocities of ITRF stations.
As a consequence, we face systematic errors involving the connection between the ICRF and ITRF realizations, which cannot be fixed by datum correction during the current solution.

\vspace{-4ex}
\section{The connection between the ICRF and ITRF}

A CRF realization obtained from a global VLBI solution depends on the tie to the ITRF.
There are several problems that affect the VLBI-derived celestial reference frame, such as dependence on the ITRF datum, the set of reference stations used by the
IVS Analysis Centers, and modeling of non-linear station motion.

The actual station movement cannot be represented to millimeter accuracy in the framework of the
ITRF using a linear drift model with occasional jumps.
Many stations show significant non-linear terms in their position time series.
The most common are the exponential movement of stations due to post-seismic relaxation and seasonal signals;  both cause a deviation from the ITRF model at the centimeter level.
This has two consequences.  First, using these stations may adversely affect the ICRF orientation.
Secondly, their use causes errors in the daily/session Earth Orientation Parameter solution.

The impact of the ICRF on the ITRF seems not to be fully understood yet.  Possible mechanisms include the nature of the global VLBI solution.

\vspace{-3ex}
\section{The ICRF status and prospects}

The current ICRF realization, ICRF2, created in 2009, provides much improvement over the first ICRF in terms of the total number of sources, the
source position precision and accuracy, and the stability of the axes.
However, there are severe problems preventing further ICRF improvement, especially with respect to systematic errors:
\begin{itemize}
\itemsep = 0.0ex
\item Uncertainty in ICRS definition at the $\mu$as level;
\item Uneven distribution of sources and source position errors over the sky (only about 2.5\% of the
observations are in the declination band $-90^o$ to $-30^o$);
\item Proper (physical) and apparent (instrumental and analysis) source motions;
\item Source structure and its variability; and
\item Dependence of source positions on wavelength (analogous to the color equation in optical astrometry), astronomical and geophysical models, observing network, and analysis strategy.
\end{itemize}

The ICRF history shows that successive versions were issued at intervals of about five years.
It should therefore be reasonable to set a goal of completing the next ICRF versions in 2014 and 2019.
It should be mentioned that the ICRF, ICRF-Ext.~1, ICRF-Ext.~2 represent, in fact, the same system based on the unchanged coordinates of the 212 defining sources.
Preserving the positions of the defining (and, as a rule, most observed) sources, originally computed in 1995, may be the main reason for the ICRF systematic errors at a level of 0.2--0.25~mas.
On the contrary, all the ICRF2 source positions were adjusted in a single global solution independently, and are tied to the ICRF only by the overall orientation (NNR constraint).
One of several options would be to keep this strategy for the next realizations, and name them ICRF3 and ICRF4.  In this regard, perhaps it would be better to use the term ``core sources'' instead of ``defining sources'', analogous to the ITRF core sources.

Two important motivations for the creating the next ICRF version in a few years are:
\begin{itemize}
\itemsep = 0.0ex
\item It is expected that there will be about 9--10 million VLBI observations by 2014--2015, 1.5~times
the number used for ICRF2.  This 
will allow a substantial improvement in the position of many (now) poorly observed ICRF sources.
\item Comparison of the ICRF2 with the latest CRF catalogues indicates that the ICRF2 may be affected by systematic errors at a level of 20--30 $\mu$as.
\end{itemize}

Of course, preparation for the next ICRF catalogue should involve an intensification of the observations of new and poorly observed southern sources.

\vspace{-2ex}
\section{Conclusion}

In addition to the efforts of the IERS and technical services to improve the CRF and TRF, the following steps seem to be urgent.
\begin{enumerate}
\itemsep = 0.0ex
\item[1.] For a timely ICRF update, it is important to establish continuous monitoring of the ICRF systematic errors by comparison to the latest CRF realizations.
\item[2.] We must identify new core ICRF sources in the southern hemisphere and start
observations of them, along with sources that were poorly observed in the ICRF2.
A practical way to do it would be to include more ICRF sources in the regular IVS sessions, such as R1 and R4.  A trade-off between the practically insignificant degradation of the EOP precision and the
long-term improvement of the ICRF can be easily found.
\item[3.] A method of describing and predicting the non-linear motion of TRF (and CRF?) 
objects at the $\mu$as/mm level of accuracy is needed.
\item[4.] An agreement on a standard set of VTRF (IVS TRF realization) core stations is needed.
\end{enumerate}

\end{document}